\documentclass[letterpaper, 10 pt, conference]{ieeeconf}  

\IEEEoverridecommandlockouts                              

\usepackage{soul}
\usepackage{booktabs}
\usepackage{subcaption}
\usepackage{graphicx}
\title{\LARGE \bf
Observations of atypical users from a pilot deployment of a public-space social robot in a church
}

\author{Andrew Blair$^{1}$, Peggy Gregory$^{1}$ and Mary Ellen Foster$^{1}$
\thanks{$^{1}$Andrew Blair, Peggy Gregory and Mary Ellen Foster are with the School of Computing Science,
        University of Glasgow, United Kingdom. Correspondence address:
        {\tt\small andrew.blair@glasgow.ac.uk}}%
}

\begin{document}

\maketitle
\thispagestyle{empty}
\pagestyle{empty}

\begin{abstract}

Though a goal of HRI is the natural integration of social robots into everyday public spaces, real-world studies still occur mostly within controlled environments with predetermined participants. True public spaces present an environment which is largely unconstrained and unpredictable, frequented by a diverse range of people whose goals can often conflict with those of the robot. When combined with the general unfamiliarity most people have with social robots, this leads to unexpected human-robot interactions in these public spaces that are rarely discussed or detected in other contexts. In this paper, we describe atypical users we observed interacting with our robot, and those who did not, during a three-day pilot deployment within a large working church and visitor attraction. We then discuss theoretical future advances in the field that could address these challenges, as well as immediate practical mitigations and strategies to help improve public space human-robot interactions in the present. This work contributes empirical insights into the dynamics of human–robot interaction in public environments and offers actionable guidance for more effective future deployments for social robot designers.
\end{abstract}

\section{Introduction}
Within a lab environment, or a controlled real-world environment such as a longitudinal study, the motivations or goals of a user can be predicted with a great degree of confidence. They will have likely seen a recruitment poster or email, been given a participant information sheet about what is going to happen during the experiment or study, and crucially agreed \textit{upfront} to interacting with the robot. Examples of this can be seen with a home companion robot designed to help children improve their reading \cite{zhao_lets_2022} and a mental wellbeing robot for the workplace \cite{axelsson_oh_2024}. They can often also be given training or supported by the researchers prior during their first interaction \cite{cagiltay_toward_2024}. In longitudinal studies, the user is afforded the opportunity to build and develop mental models of how robot interactions operate over multiple sessions.

However, in public spaces, all of the above cannot be guaranteed. As a starting point, the user almost certainly would not have expected to encounter the robot given currently, as Zawieska states, robots are \textit{``evidently absent in everyday environments''} \cite{10.1145/3695772}. They are not likely not afforded the opportunity for training on how to interact with the robot, as the focus is often on short-term interactions and the ability for the robot to act autonomously and without human intervention is a key component of any real-world study. 

Mintrom defines public spaces as \textit{``anywhere that groups of people who may or may not know each other can freely assemble, move about, and interact''} \cite{mintrom_robots_2022}. This presents one of the most complex socio-technical landscapes for a robot to navigate and perform in, and therefore to design for. While the robot's task can be co-designed with the organisation or space where the robot has been placed, the ability to accurately cater for this exceptionally diverse group of people is extremely difficult. 

To help further the design of public-space social robots, we present in this paper a non-exhaustive list of user encounters during a recent 3 day pilot deployment of our social robot within a church. We then discuss possible resolutions and mitigation strategies to help improve the robustness of robots deployed within real-world public spaces.

\section{Background}
We partnered with a large working church and visitor attraction in the United Kingdom to explore the potential roles of a social robot within the church. The church attracts a wide range of people, with some viewing the space as a centuries-old spiritual sanctuary and others seeing it as a place of architectural beauty with great photo opportunities. They have previously explored other technologies to help increase visitor engagement and to augment the visitor experience, and were now interested in exploring the use of social robots.

After an extensive co-design process \cite{blair_blessing_2025}, we developed a first version of our robot system to pilot within the church. The scope of the robot was to conduct an exit survey with visitors, and for visitors to be able to ask questions about the history of the church. The robot was also multilingual, being able to understand and respond in English, French, German, Italian, Spanish and Mandarin.

\begin{figure}[h]
    \centering
    \includegraphics[width=0.8\linewidth]{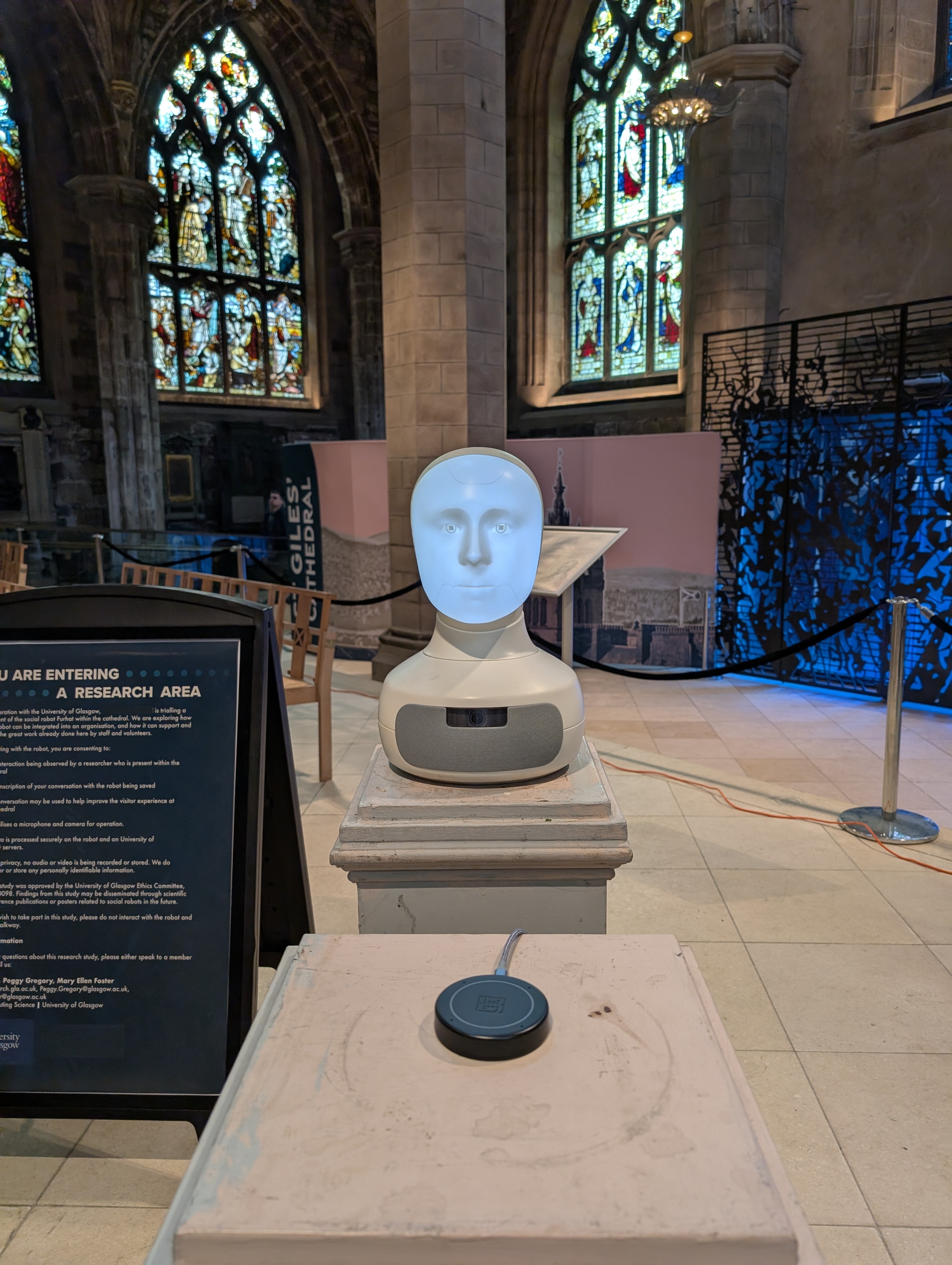}
    \caption{Furhat robot deployed in the church}
    \label{fig:setup}
\end{figure}

We used the Furhat robot \cite{al_moubayed_furhat_2012} with an external beam-forming microphone placed directly in front of the robot. This was connected to an external computer; this was to overcome issues with the automatic gain control of the microphone when the robot was speaking. The back-projected image was set to the ``Titan'' profile, which presents a robotic face to the user. We used text-to-speech from Amazon Polly, and paired voices with feminine features to the user selected language. A visual representation of the system can be seen in Figure \ref{fig:setup}.

\section{Atypical users we encountered}
The subsequent described atypical user encounters were observed by the first author during the pilot deployment. The observations were recorded through both contemporaneous and post-hoc field notes. We declined to use video recordings for ethical and consent reasons, as we felt users having to provide informed consent would alter the dynamics and produce an unnatural interaction. The first author was always present on the church floor, but not situated next to the robot; instead, a dynamic, roaming approach was employed to observe users. 
 \subsection{The Phubber} 
 This user is often willing to actually engage with the robot, but with one key, non negotiable understanding --- they must be able to record the interaction with their mobile phone. This likely stems from the lack of robot presence in most people's daily lives \cite{10.1145/3695772}, and the resultant novelty factor the robot represents to a person \cite{reimann_social_2024}.

\begin{figure}[h]
     \centering
     \includegraphics[width=1\linewidth]{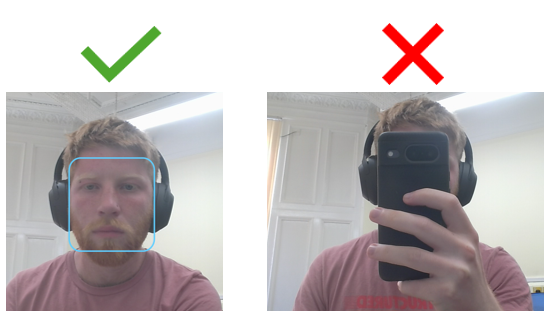}
     \caption{Example of how objects can obscure visual sensing. Reproduced in laboratory.}
     \label{fig:obscured}
 \end{figure}

 Unfortunately, this introduces a problem for the robot -- one of it's primary sensors of human social signals, visual, has now been obfuscated by the presence of the mobile phone. Figure \ref{fig:obscured} shows an example of what the robot can observe when a user is using their mobile phone to record their interaction versus without. We can see that facial landmarks can be almost completely obscured. This problem has also been faced with the use of medical face masks; however, the eyes are almost always present, which with our mobile phone user we can also lose. In effect, we can see the user is partially ``phubbing'' the robot: ``ignor[ing] someone you are with and giv[ing] attention to your mobile phone instead.'' \cite{cambridge_dictionary_phubber_2025}.

 \subsection{The Starer}

 This user will approach the robot. We take a view of this user from that of a ``proactive robot''; that is, the robot will speak first based on a social cue; a common indicator in both humans and robots that someone wishes to speak with you is gaze. The robot will then generate an utterance, attempting to start a conversation with the user. However the user will not respond at all. They will simply continue to look at the robot, and then eventually walk off and continue their day. This can result in the classical scenario of the robot ``talking into the void''; robots often relying on gaze and proximity as embodied cues of communicative intent, and this user effectively breaches the often times fragile interactional order that exists in human-robot interactions.

 \subsection{The Reader}
Informed consent is one of the key aspects of any research project. However, when attempting to conduct real-world research with people, it is often impractical or unnatural to offer a standard written consent form due to time constraints or because it fundamentally affects what is being studied \cite{huber_informed_2021}. To address this in our study, we placed a large A2 information sign next to the robot, explaining our ethnographical approach and what data we were collecting, as well as that all our sensor inputs were processed locally and not via third party providers.

We observed a number of instances of people simply reading the information sign and then not interacting with the robot. They would therefore also stand in front of the robot while reading the sign, and other potential users of the robot would see this and then decide to not interact with the robot as they did not want to wait. 

On reflection, we likely made the information sign's wording too complex and used technical terms that could confuse a layperson. It also changed the dynamics of the interaction with the robot; instead of people approaching the robot, they approached the sign first; in a non-research project the sign would not exist at all.

 \subsection{The Abuser}
 Physical abuse of public-space robots is an active research area \cite{brscic_escaping_2015}, with various mitigation strategies being developed and tested to help prevent damage and undesirable interactions from human actors. Similarly, in wider conversational systems, people carry out verbal, often sexist, abuse of the system \cite{cercas_curry_metoo_2018}.

 We observed a group of visitors speaking to our robot who attempted to engage in increasingly off-topic responses. The text-to-speech voices we used were designed to imitate the features of a traditionally female voice. In consultation with our stakeholders, we implemented a feature where if the user was repeatedly not engaging in questions surrounding the cathedral it would end the interaction. However, when the robot stopped the conversation, this led one of the group to call the robot ``a whore''. This was particularly unexpected given our deployment context; the church is usually considered a space where people show respect to the people and objects within it, but it could be considered that a robot does not fit into these revered categories.
 

 \subsection{The Crowd}
 \begin{figure}[h]
     \centering
     \includegraphics[width=\linewidth]{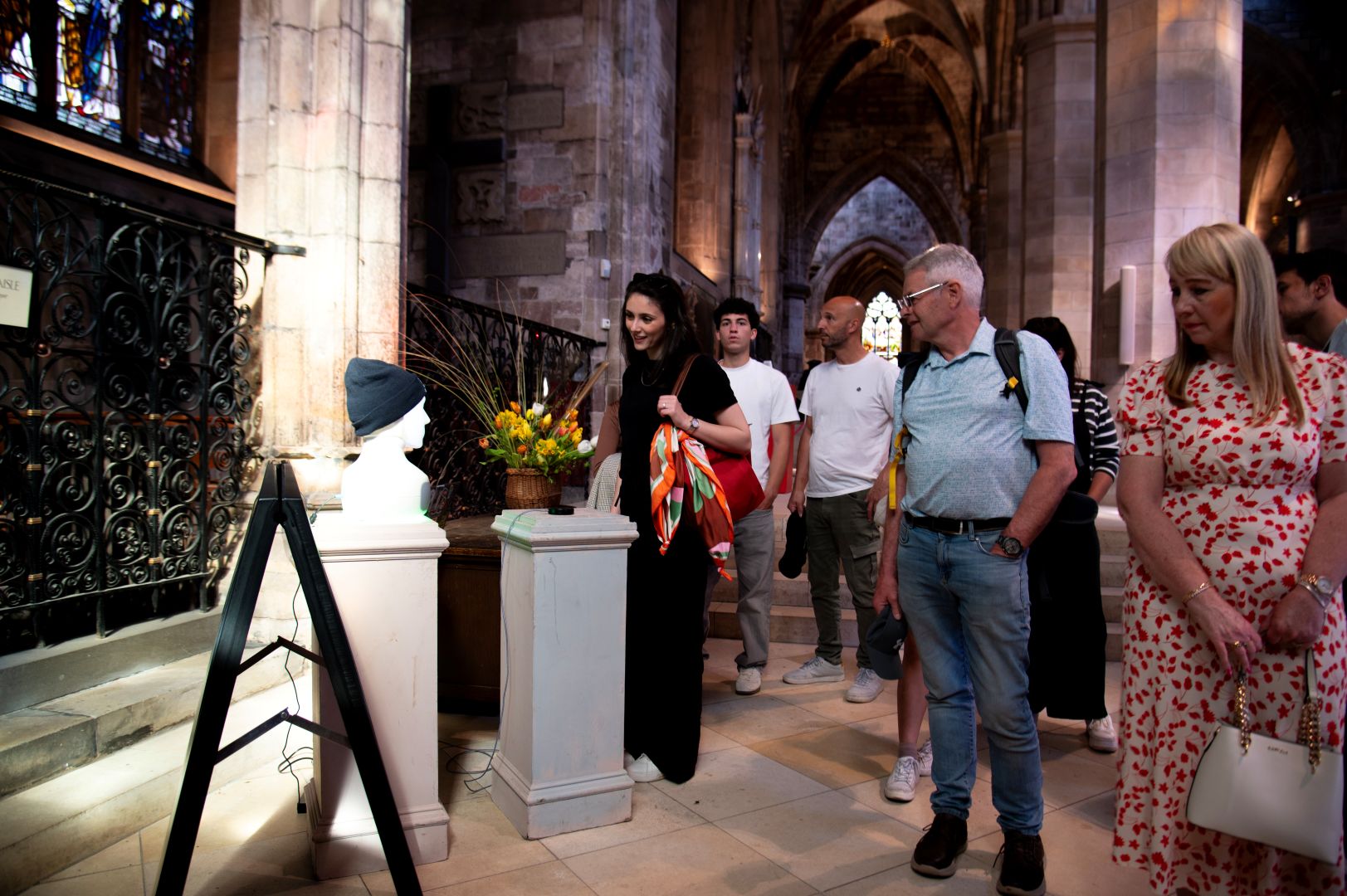}
     \caption{Crowd forming around robot, despite it being a single-party interaction.}
     \label{fig:crowd}
 \end{figure}

 While encouraging from a roboticist's point of view, crowds forming around the robot can quickly change the dynamics of the environment. We observed that when one person was talking to the robot (see Figure \ref{fig:crowd}) unrelated parties would stand and watch the interaction. This can cause knock on effects however, such as disrupting the flow of people through the space itself, and also can introduce a more complex audiovisual scene.

 \subsection{The Disinterested}
\begin{figure}[h]
    \centering
    \includegraphics[width=0.8\linewidth]{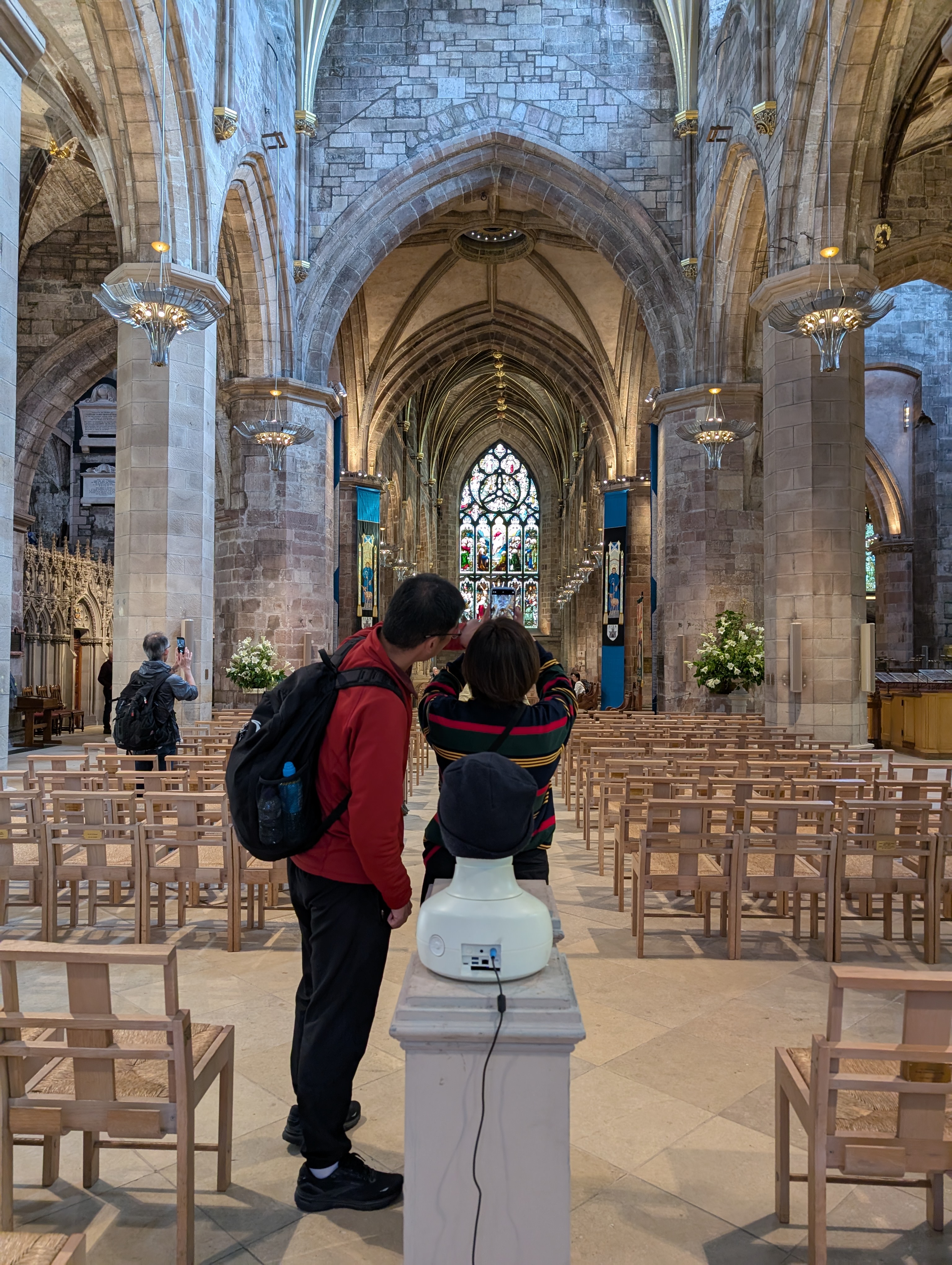}
    \caption{Couple taking photographs of church in front of the robot.}
    \label{fig:dontcare}
\end{figure}

 The typical disinterested user is well known, with them briefly glancing at it, walking by quickly or otherwise ignoring the robot.
 However, we wish to highlight another form of disinterested user, where the user can actively be hindered by the robot. Placement of non-mobile robots is often one of the most important design decisions, with researchers often attempting to make their robots as visible and obvious to potential users as possible. However, the goals of the robot can actually compete with the goals of the humans within the space. In Figure \ref{fig:dontcare}, we can see tourists attempting to photograph the church as part of their visit. They are actively leaning against the plinth the microphone is situated on. This can even go further, with visitors entering our cordoned off area such to get their photo of the church, introducing risk to both themselves and the robot due to our power supplies and cabling being contained within this area.

\subsection{The Button Pusher}
Beam-forming microphone arrays are often used in robotics to deal with noisy, reverberant environments to allow the robot to isolate user speech. An often used trick, and recommended by some robotics companies, is to place the microphone away from the robot and closer to the user. This therefore cuts the distance that the user's speech has to travel and can boost the signal-to-noise (SNR) ratio of the speech signal, resulting in more effective beam-forming, a cleaner signal, and therefore more accurate transcription. 

However, the general public does not necessarily recognise these electronic components as microphones, instead electing them to be buttons or another way in which controlling the robot. This can then end up having the opposite effect; by tapping on the casing, or picking the microphone up, this introduces severe handling noise to the resultant signal. The rotation and movement of the microphone will also degrade the effectiveness of the beam-forming process. The algorithms rely on the microphone array having a known geometry to calculate a steering vector and then perform delay and sum combination of the microphone channels; this is then disrupted by any movements of the microphone mid signal capture.

\subsection{The Robot Toucher}
Touch is one of the dominant sensory inputs we have as humans \cite{oshaughnessy_sense_1989}. 
However, when dealing with expensive and often delicate research robots, we don't always necessarily want to encourage this behaviour. For example robots like Furhat, a table-top humanoid head, relies on two servo motors contained inside the robot to control the rotation and movement. As the servo motors actively try to hold the position they are given by the robot, external pressure from a user could result in damage to these components. 

We observed many children and adults alike touching the face mask and head of the robot, despite a plinth being positioned in front of the robot that an external microphone was placed on. We did not intervene apart from in one instance where the user actively tried to roll the head of the robot. 

\subsection{The Preoccupied}
We observed many visitors to the cathedral not even notice the robots presence. Often, humanoid robots have small form factors, designed to be unintimidating and create an approachable social presence, or to reduce production costs. 

As a result, the robot can often go overlooked unintentionally by visitors. This presents a challenge in terms of proactivity; a robot that is too proactive could be perceived as annoying or obtrusive, especially within a church setting, but without it the robot may lose potential interactions. 

\subsection{The Conscientious Objector}
Manifestations of technophobia, or even \textit{robophobia}, have been described in previous studies \cite{jin_navigating_2024}. Often this can be represented as a total avoidance or apprehension to artificial intelligence. We conversed with one user who quite happily ``speaks'' to ChatGPT on their phone, and uses AI within their personal and work life. However, they do not want to be perceived as using ``AI in public''; they indicated some form of social stigma from interacting with a robot. While we note this could be partially due to our deployment context, another user mentioned they did not want the robot to start talking as they feared it could change other people's enjoyment of the space, such as disrupting the quiet of the space, and they did not want to be looked at by other people.





\section{Possible Resolutions}
\subsection{Theoretical}
To address our user encounters that encompass sensor occlusion and noise, there are a number of possible machine learning approaches that could be developed. However, these techniques rely almost exclusively on deep learning; which therefore requires significant data capture. Determining how to ethically capture such data presents a multifaceted challenge. Within public spaces, an individual can not generally expect a right to privacy and therefore recording. However, these lines start to become blurred when you begin recording in sensitive settings or more intimate interactions. For example, commercial robots such as Moxi, and self-driving cars such as that from Tesla, record by default and as such allow their models to be improved through reinforcement learning. The CEO of Diligent Robotics stated on social media they have now collected 140 Terabytes of operational data from hospital deployments of their Moxi robot \cite{diligent2025}. With the case of interaction intent, there exists \textbf{no real-world} egocentric social interaction datasets for use by the social robotics community, and only one dataset that could be used as a substitute \cite{bian_interact_2025}.  

Over time, we will see the mental models of the public surrounding robot interaction develop significantly as they are more exposed to robots in their daily lives. This in turn should help to resolve some of the aforementioned problems.
However, with no developed interaction and user experience standards for robots currently existing, as do with web interfaces, this still creates a cognitive barrier to having short-term interactions with public-space robots.

\subsection{Practical}
 While the technologies underpinning social robots, such as the introduction of end-to-end transformer-based models for automatic speech recognition \cite{radford_robust_2022} and large language models \cite{minaee_large_2024}, have allowed much more robust and reliable interactions than 5 years ago, human-robot interactions can still prove difficult in the real-world. We can see multiple user behaviours explained by the gulf of execution and evaluation \cite{norman_user_1986}. Execution is defined as the actions that must be taken in a system to achieve a goal, and evaluation is defined as the understanding of the internal state of a system; the gulf represents the challenges a user has in interacting with the system. In human-robot interaction, we can see this as a very common problem; most users will present with no existing mental models of how to interact with a robot.
 However, by using some practical interventions, and making use of existing mental models surrounding existing technologies, we suggest this gulf can be bridged in the present.

The inclusion of a tablet with the robot can provide feedback on the internal state of the robot, such as displaying a transcription of user utterances or confidence levels. It can also be used to provide an additional input modality, such as language selection \cite{ashkenazi_goes_2024}. Larger embodied robots, such as Pepper and ARI, already implement a tablet as standard. However, care needs to be taken not to make it the primary modality; Aldebaran spent considerable effort encouraging developers not to rely on the tablet on Pepper \cite{noauthor_pepper_nodate}.

To prevent external sensors such as microphones being moved, attaching, gluing or otherwise adhering it to the surface it has been placed on can ensure that the beam-forming process is accurate. 
To reduce handling noise and attempt to prevent unwanted contact, we suggest physically labelling the system. Post-hoc labelling of public-space technology is surprisingly common, as can be seen in offices with equipment such as coffee machines or turnstile entry systems, as a way to help guide user behaviours. 

In the same vein, the use of a ``fun'' information sign or other visual aid to describe the purpose of the social robot can help to bridge the gulf of execution for a user. Often, users can be unsure or even wary of what the robot is there for, and the use of a traditional form of communication tool can provide a more familiar experience and therefore cognitive ease to the user.

\section{Conclusion}
In public spaces, human goals often differ from the intrinsic goals of the robot, which can lead to unexpected human reactions and behaviours. In this paper, we reported on different users we encountered interacting with the robot, and those who did not, within a large working church and visitor attraction over the course of a 3 day pilot deployment. Due to missing collective familiarity with social robots, and the associated lack of mental models a person has with them seemingly obvious interaction patterns and paradigms are not recognised by users, leading to more difficult interactions. However, we discuss a number of simplistic interventions that could help to ease the cognitive demand people experience interacting with public-space social robots. We suggest the idea that reliance on these techniques not as a failure of the technology, but showing how short-term, unexpected human-robot interactions can benefit from utilising traditional mediums and drawing on existing mental models. We hope this paper highlights how public space differs from other real-world environments, which often function more like controlled lab environments, and how social robots within public spaces need to reflect this in their design.  





\bibliographystyle{ieeetr}
\bibliography{weirdrefs,zote}

\begin{thebibliography}{10}
\bibitem{noauthor_pepper_nodate}
``Pepper is not just a tablet — {QiSDK}.''
\end{thebibliography}

\addtolength{\textheight}{-12cm}  
\end{document}